\documentclass[apj]{emulateapj}

\usepackage{graphicx}
\usepackage{apjfonts}

\shorttitle{Brightest Cluster Galaxies in a hierarchical universe}
\shortauthors{C. Tonini et al.}
\journalinfo{ApJ submitted}

\begin{document}

\title{The evolution of Brightest Cluster Galaxies in a hierarchical universe}

\author{Chiara Tonini\altaffilmark{*}, 
Maksym Bernyk\altaffilmark{*}, 
Darren Croton\altaffilmark{*}, 
Claudia Maraston\altaffilmark{**}, 
Daniel Thomas\altaffilmark{**}}
\altaffiltext{*}{Centre for Astrophysics and Supercomputing, Swinburne University of Technology, VIC 3122, Melbourne, Australia}
\altaffiltext{**}{Institute of Cosmology and Gravitation, University of Portsmouth, PO1 3FX Portsmouth, UK}

%\altaffiltext{1}{Email: ctonini@astro.swin.edu.au}

\begin{abstract}

We investigate the evolution of Brightest Cluster Galaxies (BCGs) from redshift $z\sim1.6$ to $z=0$. We upgrade the hierarchical semi-analytic model of Croton et al. (2006) with a new spectro-photometric model that produces realistic galaxy spectra, making use of the Maraston (2005) stellar populations and a new recipe for the dust extinction. We compare the model predictions of the K-band luminosity evolution and the J-K, V-I and I-K colour evolution with a series of datasets, including Collins et al. (Nature, 2009) who argued that semi-analytic models based on the Millennium simulation cannot reproduce the red colours and high luminosity of BCGs at $z>1$. 
We show instead that the model is well in range of the observed luminosity and correctly reproduces the colour evolution of BCGs in the whole redshift range up to $z\sim1.6$. We argue that the success of the semi-analytic model is in large part due to the implementation of a more sophisticated spectro-photometric model.
An analysis of the model BCGs shows an increase in mass by a factor $2-3$ since $z\sim1$, and star formation activity down to low redshifts. While the consensus regarding BCGs is that they are passively evolving, we argue that this conclusion is affected by the degeneracy between star formation history and stellar population models used in SED-fitting, and by the inefficacy of toy-models of passive evolution to capture the complexity of real galaxies, expecially those with rich merger histories like BCGs. Following this argument, we also show that in the semi-analytic model the BCGs show a realistic mix of stellar populations, and that these stellar populations are mostly old. In addition, the age-redshift relation of the model BCGs follows that of the universe, meaning that given their merger history and star formation history, the ageing of BCGs is always dominated by the ageing of their stellar populations. In a $\Lambda$CDM universe, we define such evolution as 'passive in the hierarchical sense'.

\end{abstract}


\section{Introduction}  

Hierarchical clustering arises naturally in the favoured $\Lambda$CDM cosmology and allows us to test our theories of galaxy formation and evolution in a realistic cosmological setting. Galaxy formation models embedded in the $\Lambda$CDM cosmology have resulted in remarkable agreement with observations across a range of structures at different scales in the Universe and at different times. At the level of single galaxies however, a number of problems remain unsolved. These typically manifest most acutely for the extreme objects in the galaxy population, such as Ultra-Luminous InfraRed Galaxies, sub-mm galaxies, dwarf galaxies, and Brightest Cluster Galaxies (BCGs). Whether an excessive star-formation rate, a chemical abundance issue, or an odd combination of observed properties, these objects remain a challenge to our physical understanding of the mechanisms of galaxy formation.

When considering BCGs, it is the idea of massive, metal-rich, passively evolving objects that stretches the boundaries of the models. Such features are hard to reproduce in the context of the $\Lambda$CDM theory, where structures grow in a  bottom-up fashion, and where the biggest objects are also the last to assemble.
In addition, BCGs show signs of being quite distinct from the rest of the elliptical galaxy population in clusters (see De Lucia $\&$ Blaizot, 2007); they exhibit a very tight luminosity and colour distribution, with a relatively small scatter, a feature that suggested they could be used as 'standard candles' in cosmology (Postman $\&$ Lauer, 1995). 

Massive elliptical galaxies appear to form the bulk of their stellar mass at high redshift ($z>2$) and evolve undisturbed for most of their subsequent life. 
This information is inferred from the analysis of galaxy spectral energy distributions (SEDs), colors and luminosities. In recent papers various authors have explored the uncertainty in the mass-luminosity and mass-colour relations resulting from the choice of different stellar population models used for this kind of study
(Tonini et al. 2009, 2010, Henriques et al. 2011; see also Fontanot $\&$ Monaco, 2010).
Early assembly and subsequent passive evolution of the high-mass end of the galaxy population is a serious challenge for the hierarchical scenario; the apparent speed of mass assembly for the most massive objects is hard to accomodate on the background growth of dark matter halo structures. In fact, galaxy formation models predict a significant evolution for BCGs (and massive elliptical galaxies in general) at $z<1$ (see De Lucia $\&$ Blaizot, 2007). The bottom-up growth of structures seems unable to produce very large stellar systems at high redshifts, while the same systems continue to accrete mass and form stars at lower redshifts. 

The low-redshift star formation problem is partially alleviated with the introduction of AGN feedback, which prevents star formation even when the galaxy accretes gas, although this solution still lacks a complete physical understanding\footnote{In fact, different implementations of AGN feedback in semi-analytic models cause a divergence in the luminosity function, expecially at high redshift; see for instance Tonini et al. 2010 (Fig. 8), and Henriques et al. 2011 (Fig. 2).}. 
The high redshift, early growth of big ellipticals however seems to be a more fundamental problem, because it is more closely linked to the dark matter halo assembly rate. Simply put, at $z>1$ either there are no virialized halos big enough to host the stellar masses deduced from observations, or alternatively, there are no means of cooling enough gas with the required rates in the existing halos to produce stellar systems of that mass. 

Recently, Collins et al. (2009) pointed out the inability of the hierarchical scenario to produce galaxies like the brightest cluster galaxies (BCGs) observed at redshifts $z \geq 1$. 
The SED-fitting analysis, based on the Bruzual $\&$ Charlot (2003; BC03) stellar population models, shows that the observed BCGs at $z \geq 1$ have stellar masses estimated around $10^{12} \ M_{\odot}$ which, as they point out, are 3-4 times larger than predicted by simulations, and seem to be composed of relatively old stellar populations. The estimated ages and star formation histories indicate and early rapid formation and a fast decay, with $50\%$ of the stellar mass in place at $z=5$, and $80\%$ in place at $z=3$. The timescale of the mass assembly is thus comparable with the age of the bulk of the stellar component (2-3 Gyr), a situation closely resembling monolithic collapse. To assembly $\sim 10^{12} \ M_{\odot}$ of stars in $\sim 4$ Gyr, an average star-formation rate $SFR \sim 250 \ M_{\odot}/yr$ is needed. In order to achieve a monolithic-like stellar mass growth, the main
mechanism leading to star formation cannot be represented by mergers, since at the scales of interest it would produce shocks large enough to prevent cooling. Collins et al. (2009) argue that cold streams able to penetrate the shock-heated region are then the only viable mechanism. In addition, Collins et al. (2009) claim that, even if the hierarchical assembly of dark matter halos is right, the semi-analytic treatment of baryons is deficient in high-density environments during early assembly. Thus even if the models are consistent with observations for the evolution of luminous red galaxies, they fail in reproducing the BCG population.

What is usually omitted in such discussions is the recognition that galaxy stellar mass is, in fact, inaccessible to direct observation (unless the study is coupled with dynamics or lensing analysis). The derivation of mass, age, star formation history, metallicity and dust extinction in observed galaxies is heavily model-dependent 
and prone to degeneracies intrinsic in the SED-fitting procedure (Marchesini et al. 2009, Maraston et al. 2006, 2010, Pforr et al. 2012, Tonini et al. 2009, 2010). Although Collins et al. (2009) discuss the implications of using different stellar population models (BC03 versus M05 for instance), there are other, less obvious biases in this kind of analysis. One of such biases, which is addressed in this work, is the assumption about the star formation history of the galaxy, and the intrinsic degeneracy that exists between the star formation history and the stellar population model in use. In particular in the SED-fitting technique coupled with BC03 models, in order to fit both the colours and the K-band luminosity of observed BCGs, a passive evolution scenario for the star formation history provides the best fit, which leads to an overestimation of the stellar masses and ages
(Maraston et al. 2006, Pforr et al. 2012). 
When these results are compared with semi-analytic models, the verdict is then clear: in model BCGs there is not enough stellar mass, and it is not assembling fast enough at high redshifts.

In order to dissipate any ambiguity, and to bypass any issue related to SED-fitting, let us rephrase the problem as follows: semi-analytic models, which evolve the stellar mass following the hierarchical clustering constraints, produce BCGs that are not red enough and bright enough in the K band, and therefore do not match the actual observations. 

In these terms, the claim by Collins et al. (2009) is fair: indeed, so far semi-analytic models have been struggling with BCGs luminosity and colours.
The BCG problem could be a sign of a serious fundamental flaw in the hierarchical scenario of mass assembly, but to support this statement, we need to be absolutely certain that we understand the stellar-to-dark matter mass ratio in galaxy clusters, and the mass-to-light relations in galaxies.
To these uncertainties we must add the ones associated with the physical recipes adopted in semi-analytic models, such as the cooling and star formation-feedback loop, the treatment of environmental effects. Given all these systematics, we argue that the most probable cause of the disagreement between the hierarchical scenario and the data for the BCG population is still to be found in the details of the models we use, rather than in a fundamental pillar of our theory. 

Of particular interest for this work, the modeling of the galaxy emission in semi-analytic models is becoming more and more sophisticated, and is sensitive to physical processes that affect galaxies most effectively at high redshifts (Tonini et al. 2009, 2010, 2011).
In a recent paper (Tonini et al. 2010) we tested the ability of a semi-analytic model of galaxy formation (GalICS, Galaxies in
Cosmological Simulations, Hatton et al. 2003) to reproduce the colors and near-infrared luminosities of galaxies at the high-mass end of the stellar mass function, at redshifts $1.5<z<2.5$. The introduction of the M05 stellar population models into the hierarchical framework proved to be fundamental for successfully predict the galaxy properties. We showed that the high rest-frame K-band luminosities and red rest-frame $V-K$ colors of both passively-evolving and star-forming galaxies at $1.5<z<2.5$ can be reproduced within the range of masses and galaxy ages that are predicted by the hierarchical mass assembly. In addition, the implementation of the M05 prescription has proven to be quite successful in reproducing the high-mass end of the luminosity function at $z>2$ (Henriques et al. 2011).
The tension between model predictions and observations, regarding the near-IR mass-to-light ratio, is solved by the detailed inclusion of the TP-AGB phase of stellar evolution into the model.

In this work we extend our analysis to BCGs, in order to compare our model predictions with the observations discussed by Collins et al. (2009), and a number of other datasets that follow the photometric evolution of BCGs up to $z\sim1.6$. 
To produce our predictions we use the semi-analytic model of Croton et al. (2006). The backbone of both this model and the one used by De Lucia $\&$ Blaizot (2007) to investigate the BCG evolution is represented by the so-called 'Munich model' based on the Millennium simulation (Springel et al. 2005), while the two models differ mainly
in the adopted initial mass function (IMF) and slightly in the galaxy merger prescription. 

The main novelty introduced in this work is a new spectro-photometric model attached to the semi-analytic model, based on the M05 stellar population models, and a new treatment for dust extinction. 

It is important to state that our aim is not to produce a model ad hoc to reproduce the observations of BCGs. Rather, the sophistication we introduce in our model allows us to use it as a tool to investigate the strengths and weaknesses of our physical recipes. In particular, we do not tweak the galaxy evolution to produce more stellar mass at high redshift. We show that just by introducing a more accurate photometry the model now reproduces the colour evolution of BCGs up to $z\sim1.6$, predicts an excess of luminosity in the K band, and by taking a closer look at the evolution of model BCGs, a number of interesting questions are raised.

This paper is structured as follows: in Section 2 we describe the main features of the new model, including the recovery of the star formation histories, the stellar population model, the dust extinction, and the BCG selection. In Section 3 we present the model predictions regarding the BCG K-band luminosity and color evolution, and compare these results with data from the literature, including Collins et al. (2009). In Section 4 we analyse the model BCGs star formation histories. In Section 5 we discuss our results in the context of BCG evolution in a hierarchical framework. Finally we present our summary in Section 6. 
Throughout the paper, we adopt the following values of the cosmological parameters: $H_0=70 \ km/s/Mpc$, $\Omega_M=0.27$, $\Omega_{\Lambda}=0.73$, $w=-1$.

\section{The upgraded model}    

The upgraded version of the Croton et al. (2006) semi-analytic model used in this paper is characterized by a new approach to the production of the galaxy photometry. 
The main feature of the photometric model is the output, in the form of galaxy spectra, from the far-UV to the far-IR. The advantages of producing galaxy spectra are linked to versatility; in a phase of post-processing, the model can $1)$ produce 
light-cones and shift between absolute and apparent magnitudes without the complications introduced by K-corrections, 
$2)$ produce mock galaxy catalogues tailored to particular instruments and surveys with any set of filters, and $3)$ allow the application of a variety of dust models.

By comparison, the previous version of the model calculated galaxy luminosity from photometric tables: these represent the luminosity of single stellar population on a grid of ages and metallicity, for a fixed set of filter bands. The model used the instantaneous SFR, averaged over the simulation timestep ($\sim 300 $ Myr), to add new stellar mass to the galaxy. By keeping track of its age and metallicity, the model was able to associate to it an entry from the photometric table at all times. Notice that the resolution of the star formation history is too low to keep track of the luminosity evolution of the younger stellar populations.
The practical disadvantage of such method is obvious as well: all the parameters of the photometric model must be hardwired into the inputs, meaning that any change in the desired output (for instance different photometric bands, or AB versus Vega magnitudes), as well as any change in the dust model, require a new run of the semi-analytic model. In addition, this method produced absolute magnitudes only, while the apparent magnitudes were obtained through K-corrections, that required a (crude) modeling of the spectrum and introduced systematic errors. 

To produce realistic galaxy spectra, the model required a series of revisions, the most prominent ones being the use of more sophisticated stellar population models, and a more physically-based dust extinction recipe. This required a new technique to extract the galaxy star formation histories with a level of precision that allows to exploit the photometric model at the best of its capabilities.

The setup for the spectro-photometric model is such that, once the spectra are produced, everything else (like the choice of filters, the dust recipe, the errors associated with magnitudes, the cuts in the mock catalogues) is obtained in post-processing. A detailed description of the new setup can be found in Bernyk et al. (in prep). The new model can be accessed online to produce mock catalogues, through the Theoretical Astrophysical Observatory\footnote{$http://tao.it.swin.edu.au/$} (TAO).

\subsection{Star-formation histories}    

In order to produce realistic galaxy luminosities across a wide range of photometric bands, the mix of the stellar populations in the model galaxies must be represented with an accuracy that matches the rapid variations of the stellar population emissions in their young and intermediate ages. This is expecially crucial at redshifts $z>1$, where a substantial amount of the stellar mass is formed, and where the average age of the stellar populations is significantly lower. 
As a stellar population ages, its evolution slows down, and the precision of the output star-formation history can decrease accordingly. 

In practical terms, for every model galaxy we must follow the evolution of each new stellar population generated in every episode of star formation in the galaxy merger tree. Modern stellar population models like the M05 produce spectra with an age accuracy of 1 Myr for the youngest populations, and this must be matched by the star-formation history. The star-formation history output can be achieved with an inhomogeneous grid in lookback time attached to every object in the simulation, that records the quick evolution of the newly-formed stellar populations while progressively decreasing its precision for older populations. The grid is matched exactly to the output of the stellar population models in use. 

Complicating the calculation is the fact that the internal time grid of the semi-analytic model is differently spaced, 
so that the instantaneous star formation rates are recorded every $\sim 300$ Myr. This timescale is too long to produce a realistic UV and blue part of the galaxy spectrum. But to refine the semi-analytic model timegrid down to steps of 1 Myr would be impossible in terms of data storage and running time. On the other hand, the $\sim 300$ Myr timescale is too short for the slow evolution of the old stellar populations, and it would provide an unnecessarily high level of detail. 
To overcome this issue, we need to take the information about the star formation rate stored in the semi-analytic model timegrid of $\sim 300$ Myr and pour it into the inhomogeneous star formation history grid, somehow making up for the lack of information on very short timescales.

The star formation burst has a certain shape in time, of which we know the integral every $\sim 300$ Myr. We assume that its functional form is a constant in each $\sim 300$ Myr interval, so that to transfer between the two different timegrids, we just need to spread the mass produced in $\sim 300$ Myr over the finer SFH time bins, by weighting it with the varying width of the bin. 
The bursts of star formation produces in mergers are instead recorded on a separate timegrid, and are added to the star-formation history at the moment of the event. 
Every galaxy at every redshift features a star-formation history with the same lookback-time grid, in order to insure a consistent level of precision.

\subsection{Stellar population models}   
 
Following Tonini et al. (2009, 2010, 2011) and Henriques et al. (2011), we use the M05 stellar population models to build the galaxy emission. For the purpose of the current work, the interesting difference between M05 and BC03 (which was originally featured in the Munich model various implementations) is the inclusion in M05 of the TP-AGB emission, which largely dominates the near-IR luminosity (peaking at the rest-frame K band) of intermediate-age stellar populations. The effects of this emission in K and J make the use of M05 expecially interesting for the comparison with the Collins et al. (2009) near-IR data. 

The TP-AGB light lowers the mass-to-light ratio in the K-band by a factor between 2 and 10 depending on the average age of the stellar populations in the galaxy (Maraston 2005), so the BCG K-band luminosity is expected to be higher for the same stellar mass, and the $J-K$ colors are expected to be redder. This feature is decisive for a solution of the discrepancy between model and data found by Collins et al. (2009); the increased K-band luminosity and redder colors for a given stellar mass works in favour of obtaining more realistic-looking BCGs \textit{within} the structures provided by the hierarchical mass assembly. 

Recent new stellar population models like Conroy et al. (2009; see also Conroy $\&$ Gunn, 2010) are now providing their own recipe for the advanced stages of stellar evolution such as the TP-AGB, but the near-IR emission results closer to BC03 than M05. Note that the scope of this work is not to discriminate between different stellar population models, but to highlight the effect of different recipes on our physical interpretation of the evolution of BCGs.
An implementation of the Conroy et al. (2009) and other stellar population models in the semi-analytic model and in TAO will be addressed in future works.  

\subsection{Dust}   

To complete the spectro-photometric model, we provide a new recipe for the dust extinction, which is an extension of the one adopted in Tonini et al. (2010, 2011). 
The model relies on the notion that dust grains are composed of metals released into the inter-stellar medium (ISM) mainly by massive stars in the phases of planetary nebulae and supernovae Type II,
and are subsequently destroyed by ionizing photons, such as those emitted by massive stars, with a relatively short ($\sim10^7 yr$) duty cycle (a review can be found in Boulanger, 2009, Section 3).
The dust content of the ISM therefore is tightly linked with the ongoing star formation activity in the galaxy (see for instance Xiao et al. 2012, Daddi et al. 2007, Maraston et al. 2010), and specifically in our model it increases with the instantaneous star formation rate (SFR). 
This approach provides a physical interpretation of the galaxy dust content, and is in line with ongoing investigations in the field (for instance Ferrara et al. 1999, Guo $\&$ White 2008). 

{We characterise the total dust content of the galaxy with the colour excess $E(B-V)$, which represents the relative attenuation of the B-band and V-band luminosities, and we express it as a function of the galaxy SFR with the following parameterisation: 
\begin{equation}
E(B-V)=R_{dust} A \cdot \left( \frac{SFR}{SFR_0} \right)^{\gamma} +B~,
\label{colourexcess}
\end{equation}

\begin{figure} 
\includegraphics[scale=0.4]{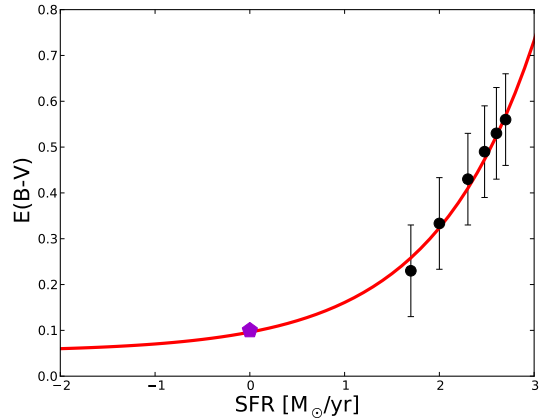}  
\caption{The relation between the colour excess $E(B-V)$ and the instantaneous star formation rate (SFR). The \textit{red line} is the relation parameterized by Equation \ref{colourexcess}, the \textit{purple symbol} represents the Andromeda galaxy, and the \textit{black points with errorbars} represent a sample from GOODS (Daddi et al. 2007).}
\label{dust}
\end{figure}

the parameters $A=(e^3-e^{-2})^{-1}$, $B=-Ae^{-2}$, $SFR_0=1.479 \ M_{\odot}/yr$, $\gamma=0.4343$ are empirically calibrated based on a set of data of the $E(B-V) \ -\  SFR$ for a GOODS sample (Daddi et al. 2007, Maraston et al. 2010) and the Andromeda galaxy. 
The dust is applied to the rest-frame galaxy spectra in the form of a Calzetti extinction curve, with amplitude $E(B-V)$. The curve is calibrated with the value $R_{dust}=3.675$ in order to have zero amplitude at the centre of the Johnson K-band (21900 Angstrom), which is generally observed to be unaffected by dust. The dust absorbs light bluewards of the K band and re-emits it at longer wavelenghts. 
Since the semi-analytic model does not provide spatial information inside single galaxies, we adopt one value of the colour excess $E(B-V)$ for each galaxy. In Fig.~(\ref{dust}) we show our E(B-V) $-$ SFR relation (\textit{red line}), and the data from the GOODS sample (\textit{black points with errorbars}) and the Andromeda galaxy (\textit{purple symbol}). 

Notice that the range of SFR values of interest in the present work is well constrained by data; even at high redshift BCGs are in general not characterised by extreme bursts of star formation. In comparison, for starbust galaxies and Ultra-Luminous IR Galaxies the determination of both the SFR and the dust extinction are far more uncertain. The present model extrapolates to the very high SFR regime rather conservatively.

\subsection{The model BCGs}  

Real-life BCGs are defined at $z=0$ as the most luminous members residing in galaxy cluster cores, usually by a selection in the K-band. 
In general they are determined to be the most massive galaxies in the clusters, bar the uncertainties in the stellar mass determination, and under the assumption of passive evolution. They are also estimated to sit at or near the minimum of the 
cluster gravitational potential 
well\footnote{Although recent works, such as Skibba et al. (2011), pointed out that this criterium can be violated in as much as ~40\% of clusters.}. 

To study the evolution of BCGs up to $z\sim1.6$, we must first characterize a galaxy cluster up to that redshift.
The definition of cluster for the local universe, based on a threshold on the estimated total mass (usually of the order of $M \sim 10^{14} \ M_{\odot}$ at $z=0$), becomes problematic and somewhat arbitrary at higher redshifts, due to uncertain evolutionary effects in the cluster mass function.
In models, a way around this problem is to define as clusters (or cluster progenitors) the $N$ most massive dark matter halos at all redshifts, and to define as BCGs their central galaxies. As discussed in De Lucia $\&$ Blaizot (2007), a choice of this sort preserves the comoving number density of BCGs at all redshifts, and our model sample is then qualitatively comparable to a sample of X-ray selected galaxy clusters down to some fixed comoving abundance. 
Even though, due to the variety of mass accretion histories for different halos, there is only a partial overlap between local BCGs and the descendants of high-redshift BCGs, the fact that these objects sit at the center of the largest dark matter perturbations at all
redshifts guarantees that they represent different evolutionary stages of a uniform class of galaxies (De Lucia $\&$ Blaizot, 2007).

In semi-analytic models the central galaxy in any dark matter halo sit at the minimum of the gravitational potential and is the most massive in that halo by definition, without any constraints on its past evolution. The other galaxies in the halo are defined as satellites, and they are subject to tidal
stripping of gas and dark matter as soon as they enter the larger halo, with the consequent truncation of the star formation and the fading of the stellar component. 
These assumptions normally imply that the central galaxy in each halo is also the most luminous in all bands, and can be defined as a BCG for cluster halos. In any case,
it has been extensively discussed in the literature (see for instance Tonini et al. 2011) that the treatment of satellites in models is too blunt, and presents an unrealistic portrait of these galaxies. For this reason, semi-analytic models are unfit to reproduce a case of a BCG that is not the central galaxy in the dark matter halo. 

We define the model BCGs as the central galaxies in the most massive $N=200$ dark matter halos for each output (i.e. redshift) of the simulation, up to $z\sim2$. 

\section{Results}  

\begin{figure*}    
\includegraphics[scale=0.9]{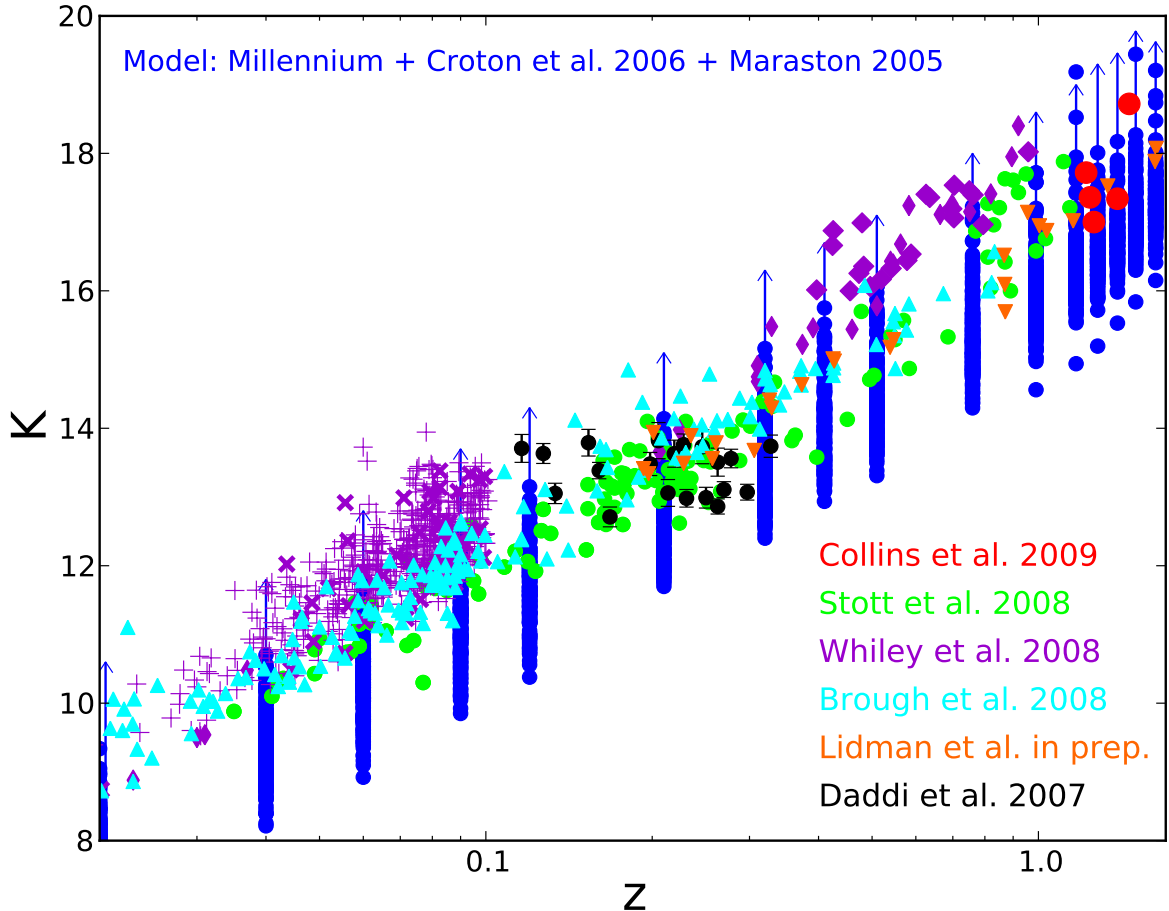} 
\caption{The observed-frame K-band luminosity evolution of the model BCGs (\textit{blue points}), compared with a number of datasets: the  sample discussed by Collins et al. (2010; \textit{red points}), two samples by Whiley et al. (2008; Hubble and SDSS data, \textit{purple diamonds and crosses}), 
a sample by Stott et al. (2008; \textit{green points}), a sample by Brough et al. (2008; \textit{cyan triangles}), a sample by Lidman et al. (\textit{orange triangles}) and a sample from GOODS (Daddi et al. 2007, \textit{black squares}).}
\label{k}
\end{figure*}

Fig. (\ref{k}) shows the evolution of the observed-frame K-band luminosity for the model BCGs (\textit{blue points}), compared with the sample discussed by Collins et al. (2010; \textit{red points}), and a number of samples from $z=0$ to $z\sim1.6$ by Whiley et al. (2008; Hubble and SDSS data, \textit{purple diamonds and crosses}), Stott et al. (2008; \textit{green points}), Brough et al. (2008; \textit{cyan triangles}), Lidman et al. (\textit{orange triangles}) and from GOODS (Daddi et al. 2007, \textit{black squares}). 

It is clear that the model BCGs lie on the bright side of the data, across the whole redshift range, and exhibit a larger scatter.
The definition of BCGs in this work implies that the observed luminosities are well in the range of the model, however the model can overshoot the brightest K-band magnitudes by about 1 mag. %%check
Note also that this effect is independent of dust, since the dust extinction curve is calibrated to be null in the K band. 

The cause of this excess luminosity lies in the adoption of the M05 stellar population models, that include the TP-AGB light,  with a caveat: if the star formation rate in the model BCGs had been null since, for example, redshift $z\sim 2$, the TP-AGB phase would not be active for a significant fraction of the stellar mass, and the K-band emission would be fainter; however, even a small star formation activity, resulting in a small percentage of intermediate-age stars that is negligible in terms of mass, produces a boost in the K-band light due to the TP-AGB phase (see Tonini et al., 2009, 2010).

This result shows that, contrary to the claim by Collins et al. (2009), the model BCGs are definitely bright enough. In other words, the stellar mass assembled by the model, constrained by the hierarchical growth of structures, is adequate, and was not the reason for the discrepancies between models and data in the past. Up to $z\sim1.6$, the extremely luminous galaxies that lie at the center of clusters are well represented in the model galaxy population. The new spectro-photometric model, quite successful in reproducing the high mass end of the luminosity function up to redshift 4 (see Henriques et al. 2011, Tonini et al. 2010), turned the discrepancy with BCGs observations upside-down. While it is reassuring to see that BCGs are no longer out of the model luminosity (and therefore mass) range, this result however highlights the degeneracy between the stellar population models and the details of the star formation history. We remind the reader that, amongst the available stellar population models, the M05-based is the one with the highest K-band luminosity.  

On the other hand, note that the inclusion of the TP-AGB emission also affects the slope of the $K-z$ relation, which is instead well reproduced. The mass-to-light ratio in the K band evolves at different speeds depending on redshift. At 
$z>1$ the bulk of the stellar mass of the model BCGs is dominated by stellar populations in the age range $1-3$ Gyr (see the next Section), therefore the K-band emission is dominated by TP-AGB stars. The K-band mass-to-light ratio of these galaxies is the lowest in the redshift range considered, and is evolving relatively fast. At lower redshifts instead the bulk of the stellar mass is composed by progressively older stellar populations, and the  
TP-AGB contribution becomes less and less important, thus the mass-to-light ratio increases and the luminosity evolution slows down. 

The fact that the model produces the correct slope of the $K-z$ relation across the whole redshift range considered shows that the model correctly reproduces the \textit{relative} evolution ot the K-band mass-to-light ratio in BCGs. With a more modest TP-AGB contribution (as is the case of the BC03 models, used in De Lucia $\&$ Blaizot (2007)), the slope of the $K-z$ relation would be steeper, and the luminosity evolution would be completely driven by the mass evolution, with a mass-to-light ratio that stays nearly constant. In such a situation, the discrepancy between the model slope and the observed one forces the conclusion that the mass evolution in the model is off.

\begin{figure*}
\includegraphics[scale=0.9]{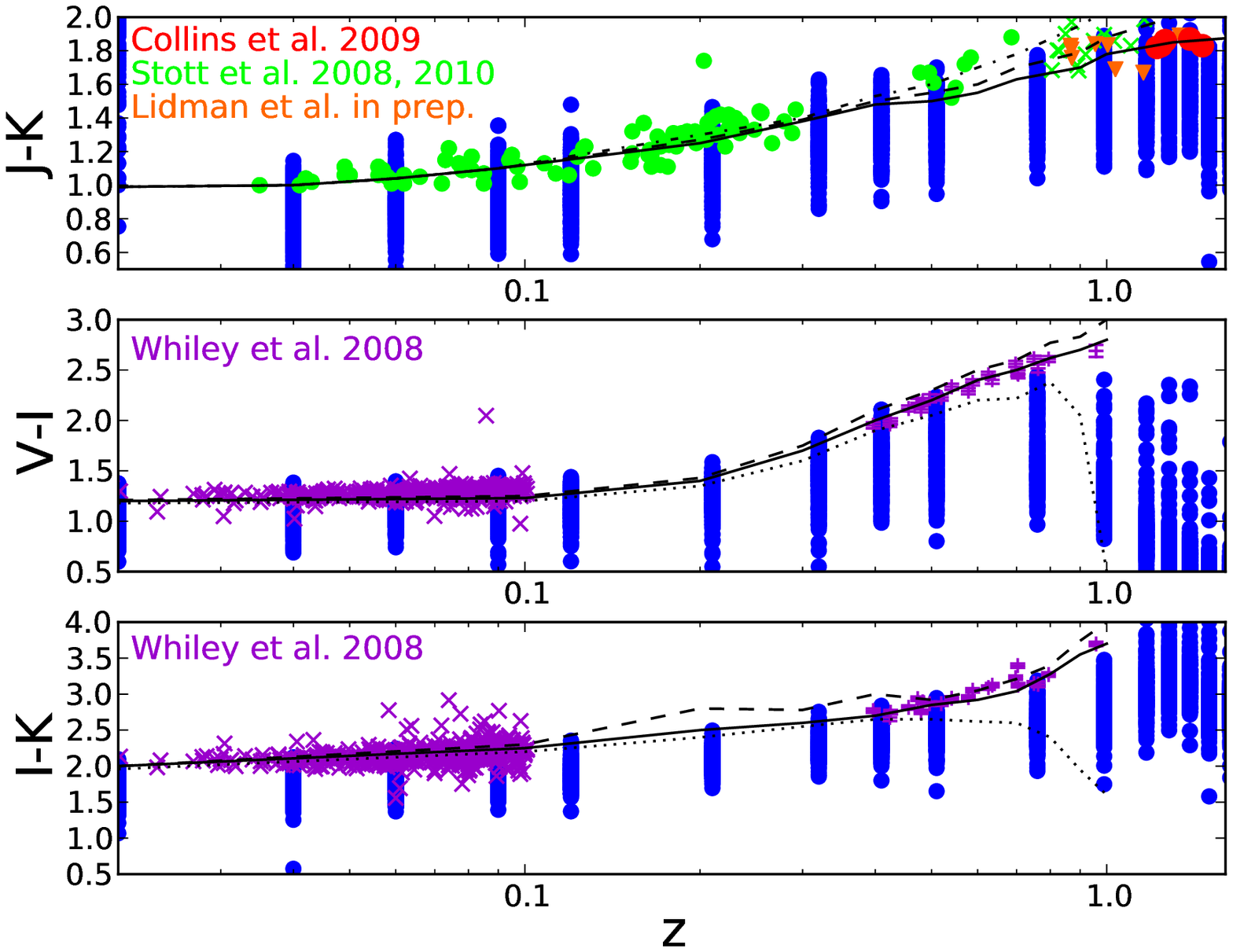} 
\caption{The color evolution of the model BCGs from redshift $z=0$ to $z\sim1.6$ (in all panels the model galaxies are represented by \textit{blue points}). \textit{Upper panel}: $J-K$, data by Collins et al. (2010; \textit{red points}), Stott et al. (2008, 2010; \textit{green points}) and Lidman et al. (\textit{orange triangles}). The lines represent toy models discussed in Collins et al. (2009) for single stellar populations (synthetised with the BC03 models), in the cases of no luminosity evolution (\textit{dot-dashed line}), with passive evolution with formation redshifts of 5 (\textit{dashed line}) and with formation redshift of 2 (\textit{solid line}).
\textit{Middle panel}: $V-I$, data by Whiley et al. (2008; \textit{purple symbols}.
\textit{Lower panel}: $I-K$, data by Whiley et al. (2008; \textit{purple symbols}. In the \textit{middle and lower panels}, the lines represent toy models discussed in Whiley et al. (2008) for single stellar populations (synthetised with the BC03 models), in the cases of passive evolution with formation redshifts of 1 (\textit{dotted line}), formation redshift of 2 (\textit{solid line}) and formation redshift of 5 (\textit{dashed line}).}
\label{jk}
\end{figure*}

Fig. (\ref{jk}) shows the predicted evolution with redshift of the observed-frame $J-K$, $V-I$ and $I-K$ colors of the model BCGs bewteen $z\sim0$ and $z\sim1.6$ 
(\textit{blue points}), compared with the observations discussed in Collins et al. (2009; \textit{red points}), the data by Stott et al. (2008, 2010; \textit{green points}), the data by Lidman et al. (in prep., \textit{orange triangles}) and the Hubble and SDSS samples by Whiley et al. (2008; \textit{purple symbols}).

The colour evolution of the model BCGs agrees remarkably well with the data. As is expected from the definition of BCGs, the data lies at the very red end of the model colour distribution, across the whole redshift range. In all colors the model is also able to reproduce the steepening of the colour-redshift relation occurring at $z\sim0.3$, but more data is needed to investigate this in more detail. More data would also be desirable to investigate the turnover of the model relation at $z>1$.

An important difference between model and data, however, is the scatter. The model BCGs exhibit a large spread in colours, shifted towards the blue. 
The cause of this is related to the model BCG evolution, and in particular their star formation history, as will be discussed in the next Section. Notice however that the current version of the model is rather conservative in the dust prescription, and the AGN feedback implementation crude (see Croton et al. 2006). Both these recipes affect the colour distribution of the model BCGs, and in particular an upgrade of the AGN feedback with more sophisticated physics is expected to reduce the colour scatter (to be addressed in future work). In the present work however, we choose not to rely too much on the still uncertain AGN feedback prescriptions to specifically reproduce the BCGs colours. 

This plot disproves the claim that hierarchical models are incapable of reproducing the very red $J-K$ colors of observed BCGs.
Notice that this result did not require a re-calibration of the model parameters. For instance, we could have raised the cooling rate or cut off AGN feedback in an attempt to correct the failed semi-analytic+BC03 galaxies, but this would have affected the entire stellar mass function, not just the BCG population. As it is, the current semi-analytic model can accommodate extreme objects like BCGs at different redshifts with a more sophisticated spectro-photometric model.

\subsection{The nature of the BCG evolution} 

At face value, this result shows that the model BCGs are realistic, albeit too bright. The slope of the colour-redshift evolution is reproduced across all colours, indicating that the model produces the correct balance between stellar populations of different ages. In addition, although the model colours tend to scatter towards the blue, the inclusion of the near-IR emission of TP-AGB stars allows the model to cover the observed range of $J-K$ and $I-K$. 

The real value in such galaxy formation models though lies not in their success or failure, but rather answering why. They allow us to take a detailed look at the galaxy population to truly test our understanding of BCGs. Specifically, how are these galaxies evolving? 
The upper panel of Fig. (\ref{jk}) shows the predicted $J-K$ redshift evolution for three toy models discussed in Collins et al. (2009): they are single stellar populations (synthetised with BC03 models) in the case of no luminosity evolution (\textit{dot-dashed line}), and passively evolving with formation redshifts of 5 (\textit{dashed line}) and with formation redshift of 2 (\textit{solid line}). The middle and lower panel of Fig. (\ref{jk}) show the predicted $V-I$ and $I-K$ redshift evolution for three toy models discussed in Whiley et al. (2008): again they are 
single stellar populations (synthetised with BC03 models), in the cases of passive evolution with formation redshift of 5 (\textit{dashed line}), formation redshift of 2 (\textit{solid line}) and formation redshift of 1 (\textit{dotted line}). 
Among the toy models, the one depicted by the \textit{solid line} seems to be most successful in reproducing the data. This model assumes that a BCG is a single stellar population which has passively evolved from $z=2$ to $z=0$, assuming the BC03 stellar population model (notice that passive evolution is the fundamental pre-requisite to treat BCGs as standard candles). 

The semi-analytic model produces the same colours. This result, albeit positive in terms of the model performance, raises much more interesting questions. 
How much do the model galaxies differ from the toy-model scenarios of passive evolution, and how accurate is such scenario for BCGs in the real universe? How degenerate is this problem, really?

\section{Star formation histories} 

As stated already, the aim of this work is not to produce a version of a semi-analytic model that reproduces a number of data samples, but rather to use the power of the model itself to push the boundaries of our understanding of real BCGs, and in particular to question the general consensus on passive evolution. With a model that matches their colour evolution and mass/luminosity range, we are in a strong position to address the pressing issue of the BCGs star formation history.

\begin{figure*} 
\includegraphics[scale=0.6]{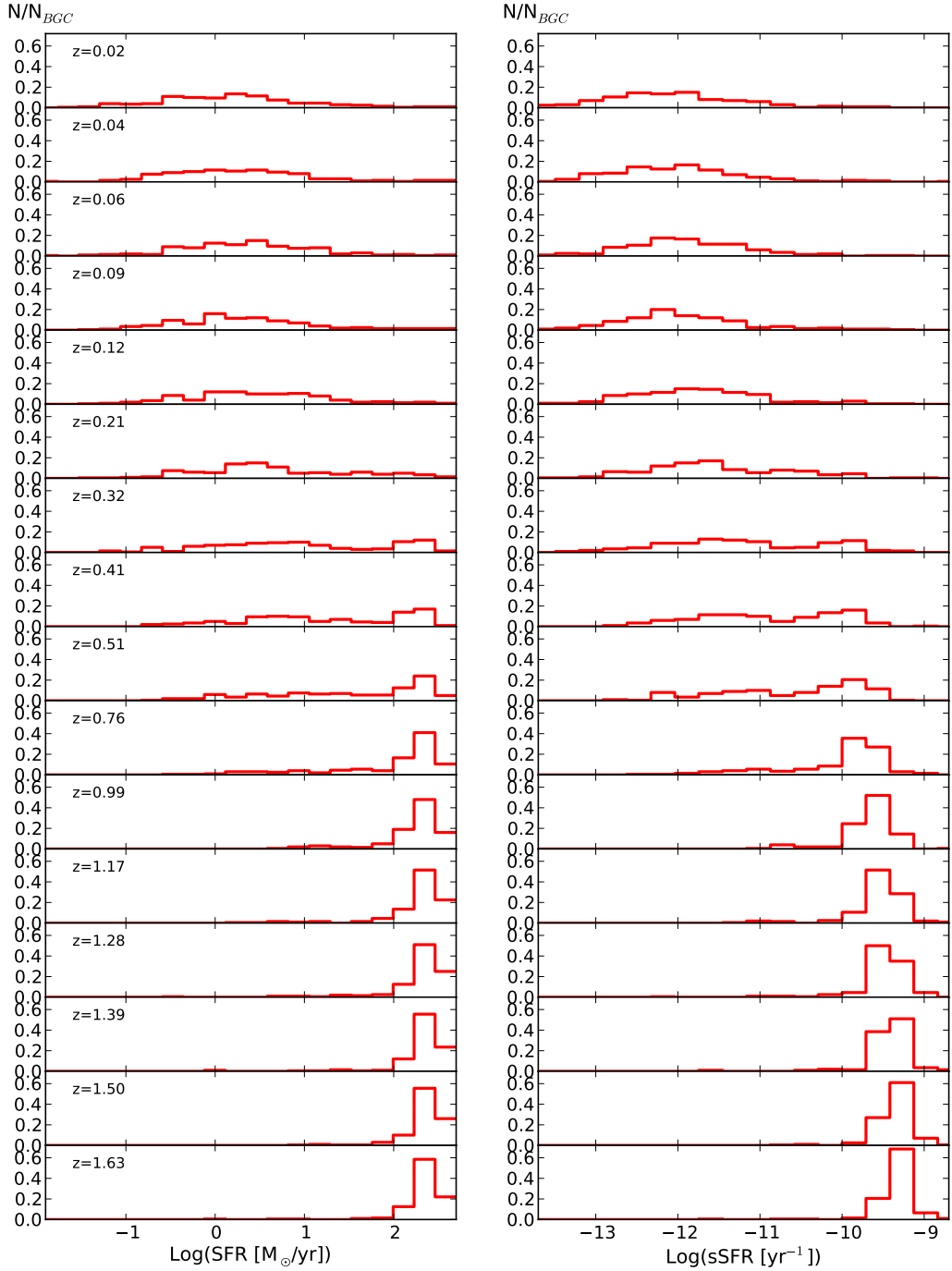}  
\caption{Star formation activity of the model BCGs from $z=0.02$ to $z=1.63$. \textit{Left column}: distribution of the instantaneous star formation rate (SFR). \textit{Right column}: distribution of the instantaneous specific star formation rate (sSFR). On the \textit{y-axis} we plot the fraction of BCGs in a given bin of star formation rate/specific star formation rate.}
\label{sfr}
\end{figure*}

\begin{figure} 
\includegraphics[scale=0.4]{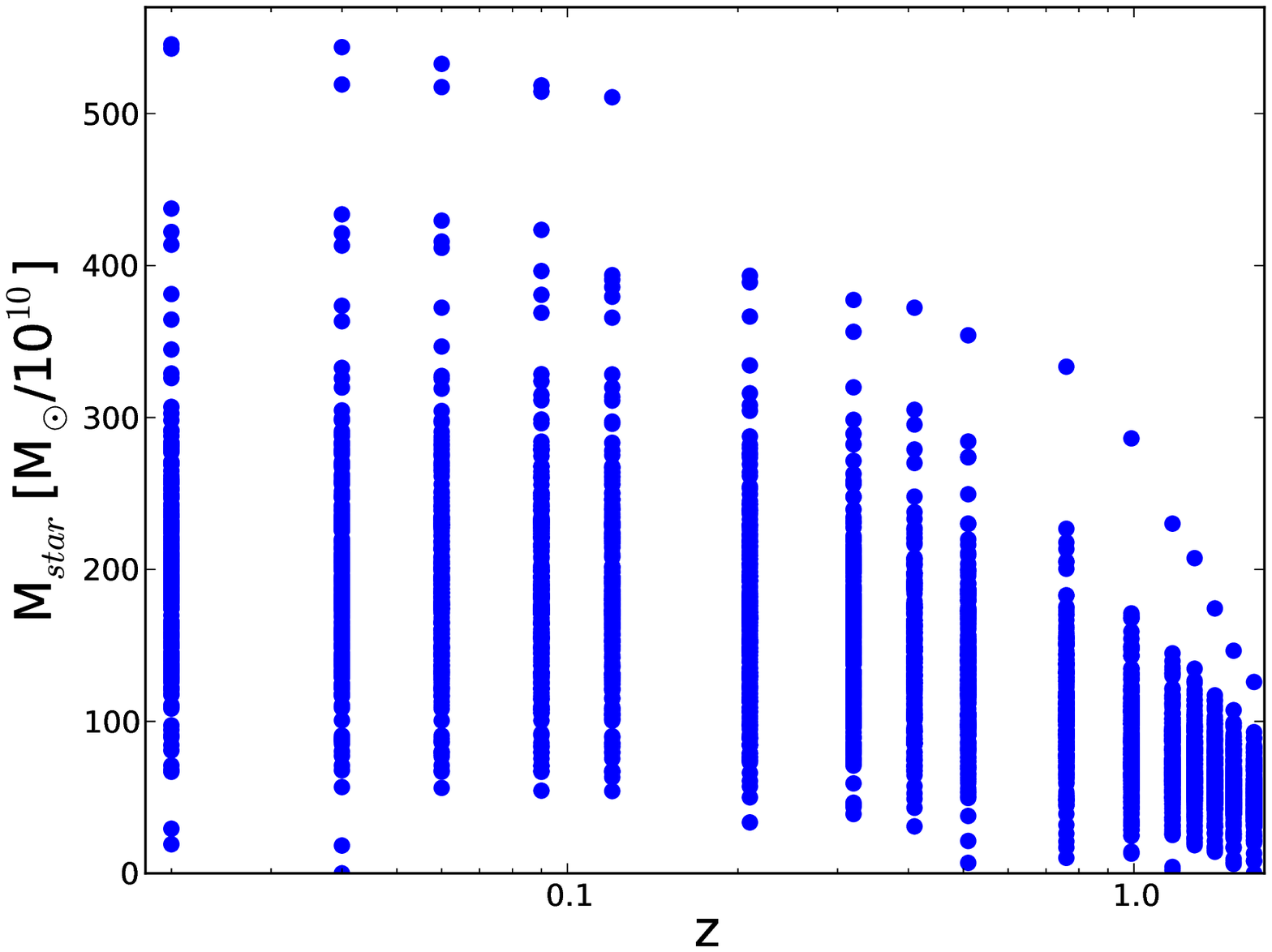}  
\caption{Stellar mass of the model BCGs (in units of $10^{10} M_{\odot}$) as a function of redshift.}
\label{mass}
\end{figure}

Figure (\ref{sfr}) shows the distributions of the \textit{instantaneous} star formation rates (SFR) (\textit{left column}) and specific star formation rate (sSFR) (\textit{right column}) of the model BCGs up to $z~1.6$. Both these quantities are defined for each galaxy as the average over the last simulation timestep, of length $\sim 300$ Myr.

At low redshifts ($z<0.2$) for the large majority of the model BCGs the star formation activity is low to moderate in absolute terms. The colours in general match the data, but a tail of bluer galaxies is produced. As clear from the values of the sSFR, this star formation activity is negligible in terms of mass growth. 
Recently, Liu et al. (2012) analysed samples of optically-selected and X-ray selected BCGs in the redshift range $z=[0.1,0.4]$, looking for signs of star formation activity. And indeed, the observed distribution of star formation rates is in accord with the results of our model, peaking around $\sim1-10 M_{\odot}/yr$.  

As redshift increases, a tail of higher SFR values starts to develop, as a result of the high-density and merging-prone environment in which these galaxies live, and by $z>0.8$ most of the model BCGs are star-forming with $SFR>100 M_{\odot}/yr$. Again, this kind of activity produces a tail of bluer galaxies, but is not enough to move the colors of the model BCGs off the data. Up to $z\sim0.8$, only a small fraction of the model BCGs is significantly increasing its mass via star formation. At $z>1$ all model BCGs are actively star forming with sSFR that indicate a significant mass growth on timescales of $\sim1$ Gyr. 

Fig.~(\ref{mass}) shows the stellar mass of the model BCGs (in units of $10^{10} M_{\odot}$) as a function of redshift, and it confirms what emerges from the star formation history. For $z>1$ the mass growth is rapid and driven by star formation. After $z\sim1$ the mass evolution slows down and stops altogether, except at the massive end of the BCG distribution where it shows a more erratic behavior. Coupled with the decrease of the sSFR values, this indicates that the mass growth at the high mass end is now driven only by mergers. 
From redshift 1.63 to redshift 0, the BCG mass growth spans a factor $2-3$, compatible with the results of De Lucia $\&$ Blaizot (2007), but the largest part of this mass evolution happens at $z>1$.

Figs.~(\ref{sfr}) and (\ref{mass}) show that the model BCGs are, in general, \emph{not} passively evolving, at odds with the toy models discussed by Collins et al. (2009) and Whiley et al. (2008), and with the diffuse consensus regarding these objects. 

\begin{figure*} 
\includegraphics[scale=0.5]{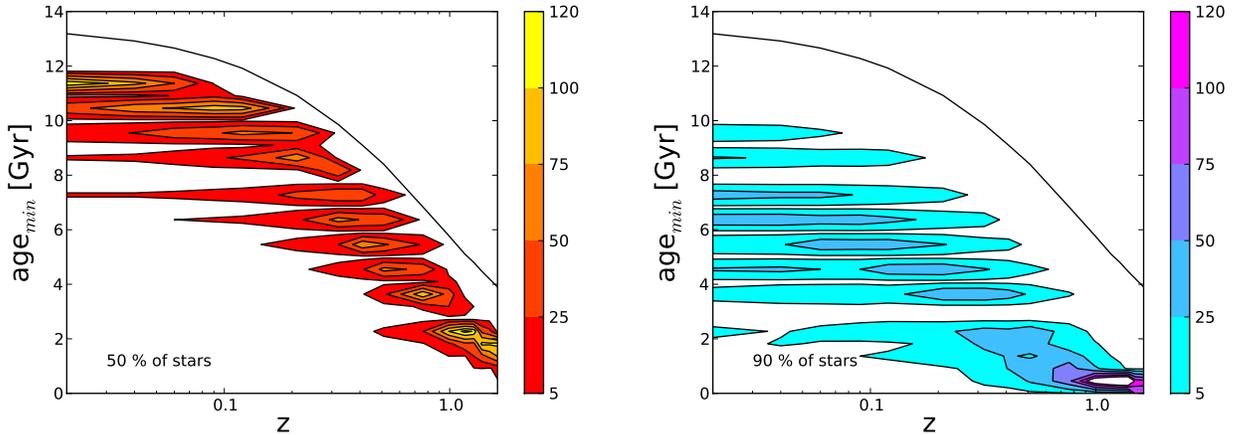}  
\caption{The distribution of ages of the model BCGs, as a function of redshift. We define the age of each BCG as the minimum age of a certain fraction of its stars, which corresponds to the lookback time of their formation. The contours represent the number of galaxies that have a certain age at a certain redshift. 
In the \textit{left panel} we define the age of the BCG based on the oldest $50\%$ of its stars, while on the \textit{right panel} we define its age based on the oldest $90\%$ of its stars. The \textit{black line} in both plots shows the age of the universe as a function of redshift.}
\label{ages}
\end{figure*}

\section{Discussion: the BCG evolution}   

A massive galaxy that is passively evolving since $z\sim2$ is exceptionally hard to produce in a hierarchical cosmology, particularly in clustered environments.
This has always been listed as a classical problem of galaxy formation in a cold dark matter universe, and up to now it seemed quite a fundamental one, with models being unable to produce enough mass (or luminosity) and reproduce the colour evolution of BCGs. However, the model considered here can reproduce the observables of BCGs and their redshift evolution, albeit with a larger scatter than the data, even within a $\Lambda$CDM cosmology. The reason is not due to an improved understanding of the fundamental physics of galaxy formation, but due to more realistic spectro-photometric modelling of the galaxy light coupled with the stochastic nature of galaxy evolution.

It is remarkable that the same BCG datasets can be reproduced with widely different evolutionary scenarios, such as passively evolving single stellar populations versus full hierarchical merger trees with stochastic star formation histories and residual star formation at low redshifts. From the observations of the K-band luminosity evolution and the colours evolution described in this work, the nature of BCG evolution is in fact inaccessible.
This should spark a healthy discussion about the degeneracies intrinsic in our modeling, including the models used on the observational side (such as the SED-fitting technique), to interpret the data and infer galaxy masses, ages, SFR, dust extinction and so on.
Recent studies (Tonini et al. 2009, Marchesini et al. 2009, 2010, Maraston et al. 2010, Pforr et al. 2012) have highlighted that the interpretation of the star formation history of galaxies from SED-fitting is heavily model-dependent, and in particular is sensitive to the stellar population models in use. 
Here lies the key to the issue: when performing the SED-fitting technique, there is a degeneracy between the stellar population model in use and the physical quantities inferred from the fit, including mass, age and the star formation history. The same is true when building a toy model like the ones discussed in the previous Section: the choice of the stellar population models determines the type of star formation history that better reproduces the data. 
In addition, these kind of toy models not only allow for a limited library of star formation histories of regular functional form, but also have no way of taking into account the accretion of stars from satellites, which is a crucial 
part of the evolution of ellipticals and BCGs in particular. So for these galaxies the toy-model approximation is especially bad, and leads to a severe misunderstanding of their evolution. 

On the other hand, the fundamental output of the model is the mass, not the luminosity. The galaxy emission is attached in post-processing to the star formation histories in the form of galaxy spectra, including the dust extinction. So there is no ambiguity about the predictions of the model regarding star formation and mass growth. 
Crucially, although the results of semi-analytic models still depend on the stellar population models of choice, they \textit{do not assume a star formation history}, nor does the star formation history come from a library of simplified theoretical functions. The model transforms gas into stars following a universal cooling recipe and the Schmidt-Kennicutt law, and the variety of outcomes is a direct consequence of environment and merger history. The model recipes for the star formation are grounded in observations, and they have a physical motivation that exists independently of the results of the model (contrary to other model parameters like for instance those connected to feedback, which are fine-tuned to reproduce the luminosity and stellar mass functions). Coupled with the fact that merger trees are completely determined by the hierarchical growth of structures in the $\Lambda$CMD cosmology, this leaves no degrees of freedom for the star formation histories. 
To force the model to produce massive passively evolving galaxies from $z\sim2$ without arbitrary fine tuning (i.e. assuming that BCGs do not evolve like normal galaxies in any respect, for example with a special AGN treatment), would require forcing the cosmology beyond the observed limits, and probably discarding the $\Lambda$CDM scenario altogether. 

The study of the evolution of BCGs make for a good example of the ambiguity introduced by our reliance on certain assumptions. 
If we assume passive evolution, then BC03 stellar population models give a good fit, but imply that the hierarchical mass assembly is wrong. If we assume BC03 stellar population models, then we have to conclude that BCGs are passively evolving in order to match their colours and luminosity, but we then have to conclude that the hierarchical mass assembly is not fast enough. If we use M05 stellar population models, then we can conclude that BCGs are not single stellar populations passively evolving but hierarchical objects with a complex assembly and star formation history, and the hierarchical mass assembly can account for such objects.

\subsection{Passive evolution in a hierarchical universe}  

The idea of passive evolution implies that no new stellar populations appear in the galaxy since the time of formation. In addition, toy models like the ones discussed by Collins et al. (2009) and Whiley et al. (2008) approximate the galaxy stellar component with a single stellar population that is left undisturbed to redden and fade. In a hierarchical universe and in a clustered environment this kind of scenario, albeit being able to fit the data, is grossly oversimplified.
In the semi-analytic model BCGs are not single stellar populations and are still evolving at low redshifts. But if we study them from the point of view of their stellar populations alone, are they young objects or are they old? How do we define the age of a realistic galaxy, composed by a mix of stellar populations, and in this case, how do we define passive evolution?  

Figure (\ref{ages}) shows the distribution of ages of the model BCGs as a function of redshift. We define the age of each BCG as the minimum age of a certain fraction of its stars, which corresponds to the lookback time of their formation. The contours represent the number of galaxies that have a certain age at a certain redshift. 
In the \textit{left panel} we define the mass-weighted age of the BCG as the lookback time by which $50\%$ of its stars were formed, while on the \textit{right panel} we define the same quantity based on $90\%$ of its stars.The \textit{black line} in both plots shows the age of the universe as a function of redshift.

This figure reveals a number of things. First, the evolution of the model BCGs fans out from $z\sim1$, showing a larger and larger variety in age as time goes by. This behaviour is driven by the differences in mass and environment, and is sustained by the ongoing star formation. This variety causes the colours of the model BCGs to scatter towards the blue.  

Second, most model BCGs are actually \textit{old}. For most of them, at any given time the majority of their stars are of an age equal at least to half the age of the universe. 

Third, the BCGs population seems to grow old following the ageing of the universe itself. There is a constant gap between the age of the oldest BCGs and the age of the universe (the black line in the plot), whether one considers the $50\%$ or the $90\%$ mark as an age indicator. We argue that this is the fastest any given galaxy can evolve, given the cosmological model. 
In fact, we could in principle have a faster ageing of the BCGs if the mergers of satellites were delayed by a few Gyrs, for example, thus making the age-redshift relation steeper. However, once we assume the WMAP cosmology $+$ $\Lambda$CDM scenario, then the hierarchical growth of dark matter structures is fully constrained, and we associate no freedom of parameters to the merger trees. This means that the times of satellite accretion are fixed. 

With this assumption in place, it is then only the star formation history $-$ in the main progenitor and in all the satellites that merge with it $-$ that determines the age of a BCG. (Notice that this makes for a consistent comparison with observations, where we can only determine the age of the stars but not the epoch of mergers).

Consequently, figure (\ref{ages}) means the following: the star formation in the model BCGs is not intense enough to offset their mass-weighted age since $z\sim 1.6$. There is not enough newly formed stellar mass to rejuvenate these objects, whose ageing is always dominated by the ageing of their stellar populations, regardless of where they were formed (main progenitor or satellites).
For comparison, a truly actively star forming galaxy (like a spiral) would show a flatter age-redshift relation, which means that the fractional increase of stellar mass is dominated by new stellar populations (and not accretion of old stars), with a rate sufficiently high to lower the mass-weighted age. 

Since the ageing of the BCG is dominated by the ageing of its stellar populations (1 Gyr every Gyr), implying that the difference between the BCG age and the age of the universe is constant, this represents an equivalent scenario to that of a single stellar population ageing in a passive evolution toy model. The difference is that we are now considering not one stellar population, but all the populations that end up forming the BCG, i.e. we are looking at the collective ageing of the stellar component in the merger tree. 
We define this behaviour as 'passive evolution in the hierarchical sense'.

\section{Summary and conclusions}   

In this paper we investigated the luminosity and colour evolution of Brightest Cluster Galaxies (BCGs) as predicted by the semi-analytic model by Croton et al. (2006). 
We upgraded the model with a new prescription for the galaxy emission, characterized by $1)$ the output of accurate star formation histories and galaxy spectra, $2)$ the use of M05 stellar population models, and $3)$ a new prescription for the dust extinction. We produced the K-band luminosity evolution and the colour evolution in J-K, V-I and I-K up to $z\sim1.6$, and compared them with a series of datasets (Collins et al. 2009, Whiley et al. 2008, Stott et al. 2008, Brough et al. 2008, Daddi et al. 2007, Lidman et al. in prep.). 

The comparison of the model BCGs with the data shows that the model can reproduce the J-K, V-I and I-K colour evolution up to $z\sim1.6$. Also, the model can reproduce the slope of the K-band luminosity evolution. The model BCGs are however too bright compared to the data, due to the coupling of the M05 photometry with residual star formation activity.

An analysis of the model BCGs, and in particular their mass assembly and star formation histories, highlights the fundamental difference between the model realisation of BCGs in the hierarchical scenario and the general consensus regarding these objects, derived from SED-fitting and toy-modeling the observations. The semi-analytic  BCGs are not passively evolving, but they increase their mass by a factor $2-3$ since $z\sim1.6$, and show signs of star formation activity down to low redshifts (although newer observations seem to support the latter feature). 
We argue that toy models are too simplistic and the method of SED-fitting plagued by the degeneracy between star formation history and stellar population models in use, which is avoided in semi-analytic models by construction.

We also emphasise that, even for semi-analytic BCGs that show evolution down to low redshifts, their stellar populations are actually old. For realistic galaxies composed of multiple stellar populations we define galaxy age based on the minimum age of an arbitrary fraction of its stars. With this metric we show that such BCGs are ageing at the same rate as the universe. We define such evolution as passive evolution in a hierarchical sense.

\section*{Acknowledgments}
CT and her co-authors wish to thank the anonimous Referee for her/his very useful comments. They also wish to thank Jeremy Mould, Alfonso Aragon-Salamanca, Chris Lidman, Sarah Brough and Diego Capozzi for the useful discussions and data exchanges. DC acknowledges the receipt of a Queen Elizabeth II Fellowship awarded by the Australian Research Council.

\end{document}